\documentclass[aps,showpacs,prl,twocolumn,amsmath,amssymb,letterpaper,floatfix]{revtex4}
\pdfoutput=1
\usepackage{graphicx}
\usepackage{tabularx}
\usepackage{wrapfig}
\usepackage{setspace}

\makeatletter
\renewcommand{\@makefntext}[1]{\parindent=1em\noindent\hbox to 1.8em{\hss$^{\@thefnmark}$}#1}
\renewcommand{\@footnotemark}{\hbox{\mathsurround=0pt$^{\@thefnmark}$}}

\makeatother

\begin{document}
% \eqsec  % uncomment this line to get equations numbered by (sec.num)
\title{$SU(2N_F)$ symmetry of QCD at high temperature and its implications
}
\author{L. Ya. Glozman}
\affiliation{Institute for Physics, 
University of Graz, Universit\"atsplatz 5, A-8010 Graz, Austria}

\begin{abstract}
If above a critical temperature  not only the $SU(N_F)_L \times SU(N_F)_R$
chiral symmetry of QCD but also the $U(1)_A$ symmetry  is restored, then
the actual symmetry of the QCD correlation functions and observables is $SU(2N_F)$. Such a symmetry prohibits existence of deconfined quarks and gluons. Hence QCD at high temperature is also in the confining regime and
elementary objects are $SU(2N_F)$ symmetric "hadrons" with not yet known
properties. 
\end{abstract}
%\pacs{11.30.Rd, 12.38.Gc, 11.25.-w}
\maketitle  
\section{Introduction}

Nonperturbatively 
QCD is defined in terms of its fundamental degrees of freedom,
quarks and gluons in Euclidean space-time. These fundamental degrees of freedom are
never observed in Minkowski space, a property of QCD which is called confinement. Only hadrons are observed.  It
is believed, however, that at high temperature QCD is in a deconfinement
regime and its fundamental degrees of freedom, quarks and gluons, are
liberated. Is it true?
Here we present results of our recent findings \cite{gg2} that suggest that this
is actually not true.

In Minkowski space-time the QCD Lagrangian in the chiral limit is invariant under the
chiral transformations,

\begin{equation}
SU(N_F)_L \times SU(N_F)_R \times U(1)_A \times U(1)_V.
\label{qcdsymm}
\end{equation}

\noindent
The axial $U(1)_A$ symmetry is broken by anomaly \cite{FU}.
The $SU(N_F)_L \times SU(N_F)_R$ symmetry is broken spontaneously
by the quark condensate in the vacuum. According to the Bancs-Casher
relation \cite{BC} the quark condensate in Minkowski space
can be expressed through a density of
the near-zero modes of the Euclidean Dirac operator,

\begin{equation}
\lim_{m \rightarrow 0} <0|\bar \Psi(x) \Psi(x)|0> = -\pi \rho(0)~.
\label{bc}
\end{equation}

\noindent
Consequently, if we remove by hands the near-zero modes of the
Dirac operator we can expect a restoration of the chiral
$SU(N_F)_L \times SU(N_F)_R$ symmetry in correlation functions.
If hadrons survive this "surgery", then the chiral partners should
become degenerate. The chiral partners of the $J=1$ mesons are shown in Fig. 1.

\begin{figure}
\centering
\includegraphics[angle=0,width=0.9\linewidth]{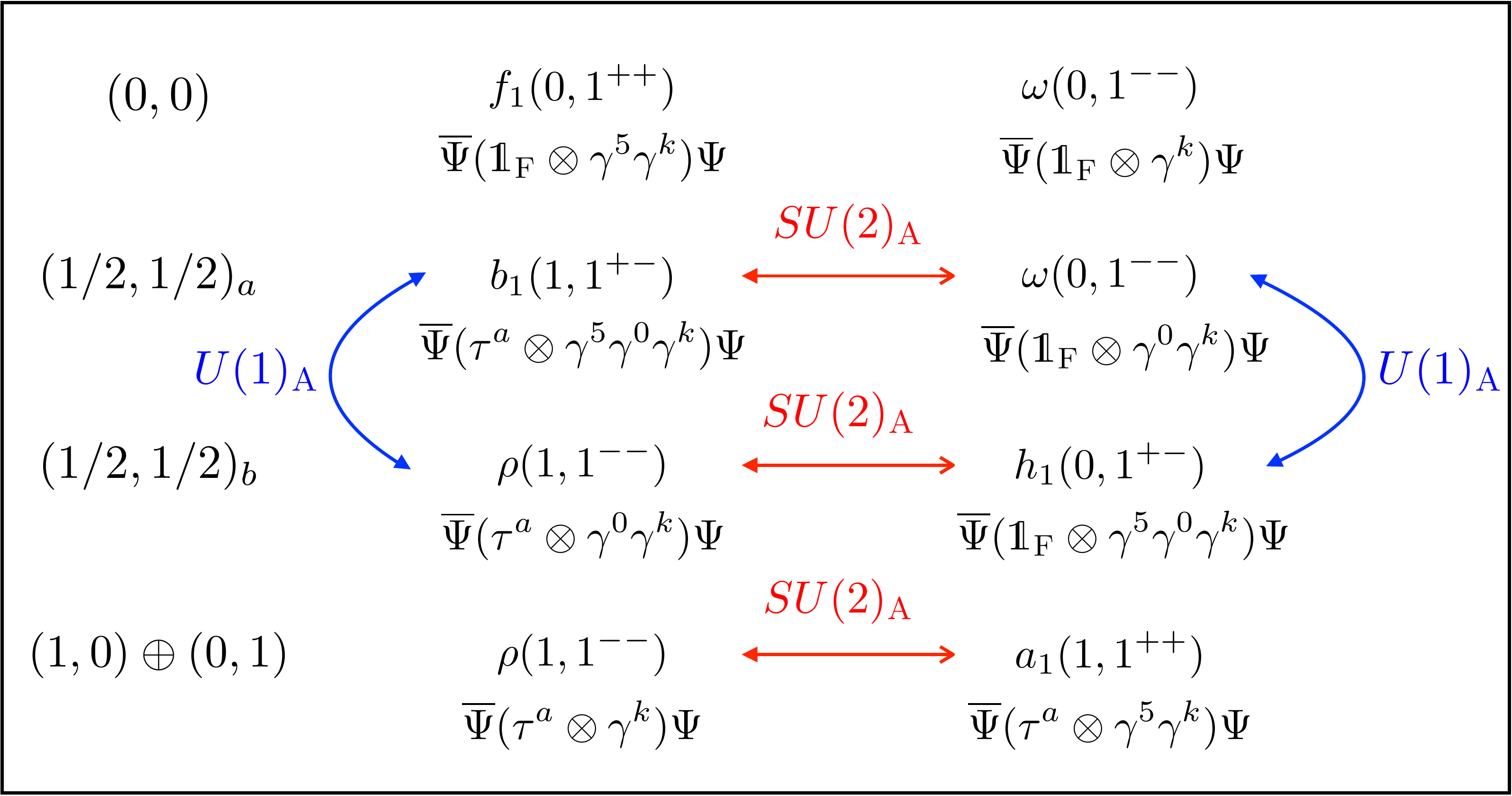}
\caption{$SU(2)_L \times SU(2)_R$ and $U(1)_A$ classification of the $J=1$
meson operators.}
\label{fig-1}
\end{figure}

It was observed in $N_F=2$ dynamical simulations with the overlap
Dirac operator that indeed  hadrons survive this truncation (except
for the ground states of $J=0$ mesons) and the chiral partners get degenerate
\cite{l1,l2,l3,l4}.
Not only the $SU(2)_L \times SU(2)_R$ restoration was observed.
Mesons that are connected by the $U(1)_A$ transformation  get also degenerate. We conclude that the same low-lying modes of the
Dirac operator are responsible for both $SU(2)_L \times SU(2)_R$ and $U(1)_A$
breakings, which is consistent with the instanton-induced mechanism for
both breakings \cite{SS}.

Restoration of the full chiral symmetry $SU(2)_L \times SU(2)_R \times U(1)_A$ of the QCD Lagrangian assumes degeneracies marked by arrows in Fig. 1.
However, a larger degeneracy that includes all possible chiral multiplets
in Fig. 1 was detected, see Fig. 2.

\begin{figure}
\centering 
\includegraphics[width=0.9\linewidth]{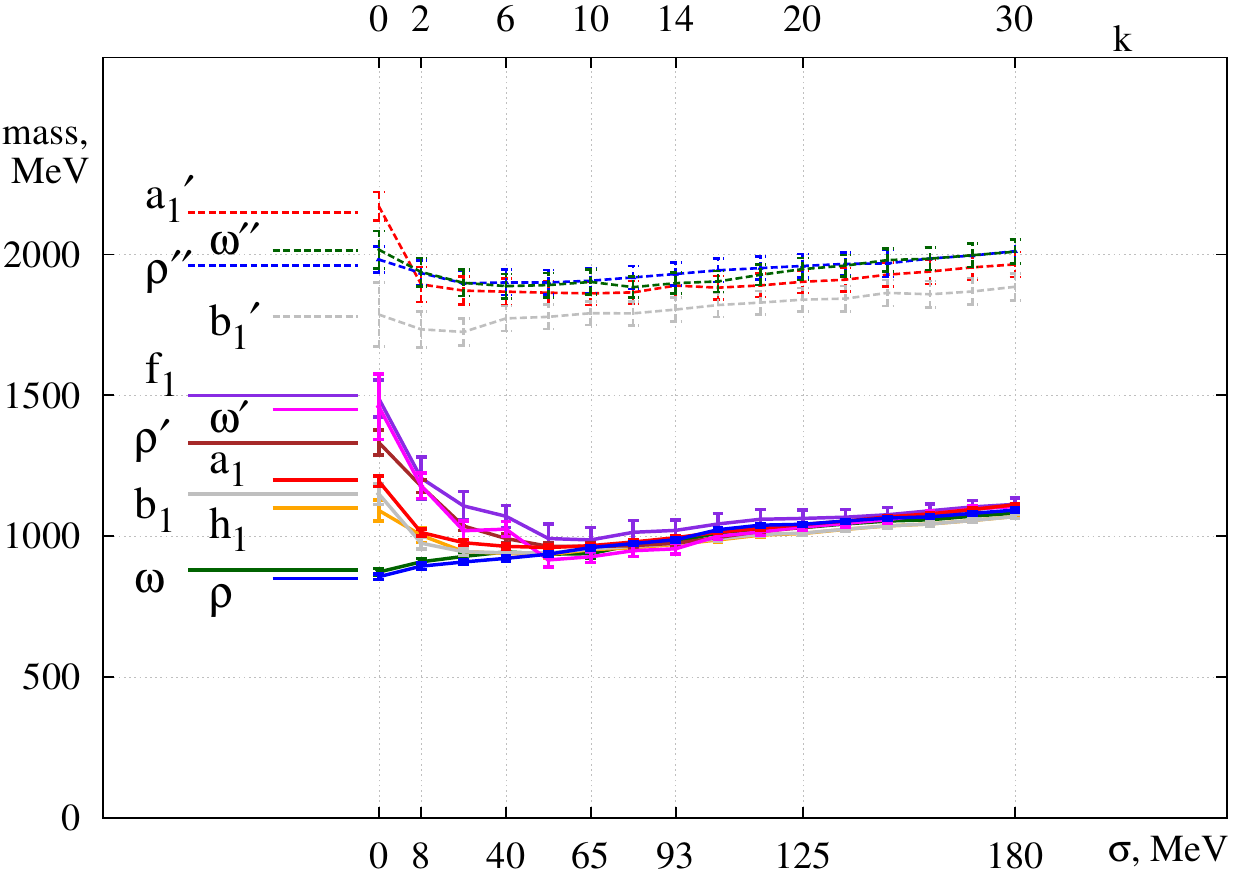}
\label{mass}
\caption{$J=1$ meson mass evolution as a function of the number $k$
of truncated lowest-lying Dirac modes.
 $\sigma$ shows energy gap in the Dirac spectrum.} 
\end{figure}
This unexpected degeneracy implies a symmetry that is larger
than the chiral symmetry of the QCD Lagrangian. This not yet known
symmetry was reconstructed in refs. \cite{g,gp} and turned out to be

\begin{equation}
SU(2N_F)  \supset SU(N_F)_L \times SU(N_F)_R \times U(1)_A.
\label{su2nf}
\end{equation}

This group includes as a subgroup the $SU(2)_{CS}$ (chiralspin) invariance.
The $SU(2)_{{CS}}$ chiralspin  generators are

$$\boldsymbol{\Sigma} = \{ \gamma^0, i \gamma^5 \gamma^0, -\gamma^5 \} \;, 
~~~~~~~~~ 
[\Sigma^i,\Sigma^j] = 2 i \epsilon^{i j k} \, \Sigma^k \; .$$

\noindent
The Dirac spinor transforms
under a global or local $SU(2)_{CS}$ transformation as
\begin{equation}
\label{V-def}
  \Psi \rightarrow  \Psi^\prime = e^{i  {\bf{\varepsilon} \cdot \bf{\Sigma}}/{2}} \Psi  \; .
\end{equation}

\noindent
The $\gamma^5$ generates a $U(1)_A$ subgroup of $SU(2)_{CS}$.
The $\gamma^0$ and $i \gamma^5 \gamma^0$ mix the left- and
right-handed components of the Dirac spinors. When we combine the
$SU(2)_{CS}$ generators with the $SU(N_F)$ generators we arrive
at the $SU(2N_F)$ group.

\section{$SU(2N_F)$ as a hidden classical symmetry of QCD \cite{gg}.}

The $SU(4)$ symmetry of $N_F=2$ Euclidean QCD was obtained in
lattice simulations. This means that this symmetry must be
encoded in the nonperturbative Euclidean formulation of QCD. Obviously
the Euclidean Lagrangian for $N_F$ degenerate quarks in a given
gauge background $A_\mu(x)$,

\begin{equation}
{\cal {L}} = \Psi^\dag(x)( \gamma_\mu D_\mu + m) \Psi(x),
\label{lag}
\end{equation}

\noindent
is not $SU(2)_{CS}$ and $SU(2N_F)$-symmetric, because the Dirac
operator does not commute with the generators of $SU(2)_{CS}$.
A fundamental dynamical reason for absence of these symmetries
are  zero modes of the Dirac operator, $\gamma_\mu D_\mu  \Psi_0(x) = 0$. The zero modes are
chiral, L or R. With a gauge configuration of a nonzero global
topological charge the number of the left-handed and right-handed
zero modes is according to the Atiyah-Singer theorem not equal.
Consequently, there is no one-to-one correspondence of the left-
and right-handed zero modes. The $SU(2)_{CS}$ chiralspin rotations
mix the left- and right-handed Dirac spinors. Such a mixing can be 
defined only if there is a one-to-one mapping of the left- and right-handed
spinors: The zero modes break the $SU(2)_{CS}$ invariance.

We can expand independent fields $\Psi(x)$ and $\Psi^\dagger(x)$  
over a complete
and orthonormal set $\Psi_n(x)$ of the eigenvalue problem 

\begin{equation}
i \gamma_\mu D_\mu  \Psi_n(x) = \lambda_n \Psi_n(x),
\end{equation}

\begin{equation}
 \Psi(x) = \sum_{n} c_n \Psi_n(x), ~~~~~ 
 \Psi^\dagger(x) = \sum_{k}\bar {c}_k \Psi^\dag_k(x),
\end{equation}
 where $\bar {c}_k,c_n$ are Grassmann numbers.
The fermionic part of the QCD partition function takes the following form

\begin{equation}
Z = \int \prod_{k,n} d\bar {c}_k dc_n  
e^{\sum_{k,n}\int d^4x 
 \bar {c}_k  c_n (\lambda_n + im) \Psi_k^\dag(x) \Psi_n(x)}. 
\label{ZZ}
\end{equation}

\noindent
In a finite volume the eigenmodes of the Dirac operator  can be separated into two classes. The exact zero modes, $\lambda =0$, and nonzero eigenmodes, $\lambda_n \neq 0$.
 It is well understood that the exact zero
modes are  irrelevant since their contributions
to the Green functions and observables vanish in the thermodynamic
limit $V \rightarrow \infty$ as $1/V$ \cite{LS,N,A}. 
Consequently, in the finite-volume calculations we can ignore  the exact zero-modes. 

Now we can  read off the symmetry properties
of the partition function (\ref{ZZ}).
For any 
$SU(2)_{CS}$ and $SU(2N_F)$ rotation  the 
$\Psi_n$ and $\Psi^\dag_k$  Dirac spinors transform as

\begin{equation}
\Psi_n \rightarrow U \Psi_n,~~~\Psi^\dag_k \rightarrow (U \Psi_k)^\dag,
 \end{equation}

\noindent
where $U$ is any  transformation  from the groups
$SU(2)_{CS}$ and $SU(2N_F)$ , $U^\dagger = U^{-1}$.
It is then clear that the exponential part of the partition function
is invariant under global and local $SU(2)_{CS}$ 
and $SU(2N_F)$ transformations, because

\begin{equation}
(U\Psi_k(x))^\dag U \Psi_n(x) = \Psi^\dag_k(x) \Psi_n(x).
\label{tt}
\end{equation}
 
\noindent
The exact zero modes  contributions 
$$\Psi^\dag_0(x) \Psi_n(x),\Psi^\dag_k(x) \Psi_0(x), \Psi^\dag_0(x) \Psi_0(x),$$ for which the equation (\ref{tt}) is not defined, are irrelevant in the thermodynamic limit and can be ignored. 
 In other words, QCD classically without the irrelevant exact zero  modes
 has  in a finite volume $V$  local $SU(2)_{CS}$ and $SU(2N_F)$ 
 symmetries. These   are hidden classical symmetries of QCD.
 
 The integration measure in the partition function
 is not invariant under
 a local $U(1)_A$ transformation \cite{FU}, which is a source of 
 the $U(1)_A$ anomaly. The
$U(1)_A$ is a subgroup of $SU(2)_{CS}$. Hence the  axial anomaly breaks 
 $SU(2)_{CS}$ and $SU(2N_F) \rightarrow SU(N_F)_L \times SU(N_F)_R$.

In the limit $V \rightarrow \infty$ the otherwise finite
lowest eigenvalues $\lambda$ condense around zero and provide
according to the Banks-Casher relation at $m \rightarrow 0$ a nonvanishing
quark condensate in Minkowski space.
The quark condensate  in Minkowski space-time breaks all $U(1)_A$, $SU(N_F)_L \times SU(N_F)_R$, $SU(2)_{CS}$
 and $SU(2N_F)$ symmetries to $SU(N_F)_V$.
In other words, the hidden classical $SU(2)_{CS}$ and $SU(2N_F)$ symmetries
are broken both by the anomaly and spontaneously.

\section{Restoration of $SU(2)_{CS}$ and $SU(2N_F)$ at high temperature \cite{gg2}} 

 Above the chiral
restoration phase transition the quark condensate vanishes. If in
addition the $U(1)_A$ symmetry is also restored \cite{A1,A2,A4} and
a gap opens in the Dirac spectrum, then
 above the critical temperture the $SU(2)_{CS}$ and $SU(2N_F)$
symmetries are manifest. The precise meaning of this statement is that the
correlation  functions and observables are $SU(2)_{CS}$ and $SU(2N_F)$ 
symmetric.

 These $SU(2)_{CS}$ and  $SU(2N_F)$ symmetries of QCD imply
 that there cannot be deconfined free quarks and gluons at any
finite temperature in Minkowski space-time. 
Indeed the Green functions and observables calculated
in terms of unconfined quarks and gluons in Minkowski space (i.e. within the perturbation theory) cannot be $SU(2)_{CS}$ and
$SU(2N_F)$ symmetric, because the chromo-magnetic
interaction necessarily breaks both symmetries.
Then it follows that above $T_c$ QCD is in a confining regime.
In contrast,  color-singlet $SU(2N_F)$-symmetric "hadrons" (with not yet known properties) 
are not prohibited by the  $SU(2N_F)$ symmetry  and can freely propagate. "Hadrons" with such a symmetry in Minkowski 
space can be
constructed \cite{Sh}.

\section{Predictions}

Restoration of the $SU(2)_{CS}$ and of $SU(2N_F)$ symmetries at
high temperatures can be tested on the lattice.

Transformation properties
of hadron operators under $SU(2)_{CS}$ and  $SU(2N_F)$ groups are given
in refs. \cite{gp,l4}. 
For example, the isovector $J=1$  operators
$\bar \Psi \vec \tau \gamma^i \Psi,(1^{--})$; 
$\bar \Psi \vec \tau \gamma^0 \gamma^i \Psi,(1^{--})$;
$\bar \Psi \vec \tau \gamma^0 \gamma^5 \gamma^i \Psi,(1^{+-})$
form an irreducible representation of $SU(2)_{CS}$. One expects that
below $T_c$ all three diagonal correlators will be different and
the off-diagonal cross-correlator of  $(1^{--})$ operators will
 not be zero. Above $T_c$ a $SU(2)_{CS}$ restoration requires that
all  diagonal correlators should become identical  and the off-diagonal correlator of
 $(1^{--})$ currents should vanish. 
A restoration of $SU(2)_{CS}$ and of $SU(N_F)_L \times SU(N_F)_R$ 
implies a restoration of $SU(2N_F)$.

A similar prediction can be made
with the baryon operators. 

We acknowledge partial support from the Austrian Science Fund (FWF)
through the grant P26627-N27.

\end{document}